\documentclass[12pt, preprint]{aastex}
\shorttitle{DUST OPTICAL PROPERTIES IN COALSACK}
\shortauthors{SUJATHA ET AL.}

\newcommand{\phunit}{photons~cm$^{-2}$~sr$^{-1}$~s$^{-1}$~\AA$^{-1}$}

\newcommand{\fuse} {{\it FUSE}}
\newcommand{\iras} {{\it IRAS}}
\newcommand{\voyager} {{\it Voyager}}

\begin{document}
\title{MEASUREMENT OF DUST OPTICAL PROPERTIES IN COALSACK}
\author{N. V. SUJATHA, JAYANT MURTHY, P. SHALIMA}
\email{sujaskm@yahoo.co.in, jmurthy@yahoo.com, shalima.p@gmail.com}
\affil{Indian Institute of Astrophysics, Koramangala, Bangalore - 560 034, India}
\and
\author{RICHARD CONN HENRY}
\email{henry@jhu.edu}
\affil{Dept. of Physics and Astronomy}
\affil{The Johns Hopkins University, Baltimore, MD 21218}

\begin{abstract}

We have used \fuse\ and \voyager\ observations of dust scattered starlight 
in the neighborhood of the Coalsack Nebula to derive the optical constants
of the dust grains. The albedo is consistent with a value of $0.28 \pm 0.04$ and the
phase function asymmetry factor with a value of $0.61 \pm 0.07$
throughout the spectral range from 900 -- 1200~\AA, in agreement with previous 
determinations as well as theoretical predictions. We have now observed 
{\it two}~regions 
(Ophiuchus and Coalsack) with intense diffuse background radiation and in 
both cases have found that the emission is due to light from nearby hot stars
scattered by a relatively thin foreground cloud, with negligible contribution 
from the background molecular cloud.

\end{abstract}


\section{INTRODUCTION}
It has long been assumed that the diffuse far-ultraviolet (FUV) background 
should be correlated with the amount of H {\small I} in the line of sight 
\citep[e.g.,][]{Mau80}. While this may be true at high galactic latitudes where 
\citet{Haikala95} and \citet{Sch} have both found the UV scattered light to be 
correlated with the 100 \micron\ emission observed using the {\it Infrared 
Astronomical 
Satellite} (\iras), albeit with different correlation factors, it is now 
becoming apparent that local effects, such as the proximity of dust to hot 
stars, can also play an important role in the level of the diffuse UV 
background \citep{Mu04,Edelstein}. This interdependence is even more apparent 
in the LMC where \citet{Cole99} have found that neither bright stars nor dust 
are sufficient in themselves to produce scattered emission; only when both are 
present with a favourable geometry is scattered light seen. In our own Galaxy, 
\citet{Lee06} found the scattered radiation in Taurus to be actually 
anti-correlated with the gas column density suggesting that the source of the 
radiation is behind the molecular cloud.

In addition to characterizing and understanding the diffuse radiation field, 
one of our scientific goals has been to extract the optical constants -- the 
albedo ({\it $a$}) and phase function asymmetry factor ({\it $g$}) -- of the 
interstellar dust grains. This has been complicated by the faintness of the 
signal and lack of knowledge about the scattering geometry 
\citep{Mathis02}. Thus, although we have observed targets over the entire sky 
\citep{Mu99, Mu04}, we have chosen to begin our modeling with two regions 
where the signal is bright and the interstellar dust distribution, from 
whence the scattering comes, is well characterized. The first of these was in 
the constellation of Ophiuchus \citep{Suj05} and the second, which we present 
here, is near the Coalsack Nebula.

\citet{Mu94} found from observations made with the two {\it Voyager} 
Ultraviolet Spectrographs (UVS) that the Coalsack was one of the brightest 
regions of 
diffuse UV emission in the sky and they attributed this emission to forward 
scattering from a relatively thin H {\small I} cloud in front of the Coalsack 
molecular cloud, a conclusion later confirmed by \citet{Sh04}. In order to 
supplement these observations, we searched for further observations made with 
the {\it Far Ultraviolet Spectroscopic Explorer} (\fuse) finding an additional 
29 observations of 21 targets, including 3 observations that were made as part 
of our own \fuse\ guest investigator observing program.

\section{OBSERVATIONS}
We have collected 34 observations (29 from \fuse\ and 5 from the {\it Voyager} UVS) 
of the diffuse radiation in and around the Coalsack Nebula (Table \ref{voy_res}). 
Of the five observations made with the \voyager~UVS, four have already been 
discussed by \citet{Mu94} and a full description of the instrument and diffuse 
observations made with it has been given by \citet{Mu99} and references 
therein. Briefly, the \voyager~UVS observe diffuse radiation from 500 - 
1600~\AA\ with a resolution of about 38~\AA. The field of view is large 
(0.1$\degr~\times$~0.87$\degr$) and integration times are long resulting in a 
sensitivity to diffuse radiation of better than 100~\phunit.

The remaining 29 observations were made with the LWRS 
($30\arcsec \times 30\arcsec$) aperture on the \fuse\ spacecraft. The four 
\fuse\ spectrographs cover the wavelength region from 850 - 1167~\AA\ with a 
resolution ($\lambda/\Delta \lambda$) of about 20000. Although intended for 
observations of point sources \citep[see][~for a description of the spacecraft 
and mission]{Moos,Sahnow2000}, \citet{Mu04} have shown that background levels 
of 2000~\phunit\ are detectable with the LWRS aperture.  

Following \citet{Mu04}, we binned the data into broad bands of about 50~\AA\ in 
width in order to increase the signal-to-noise ratio. This yields a total of 6 
independent bands (Table \ref{bands}) with sufficient sensitivity to detect the 
diffuse radiation from the Coalsack. Because the 2A2 and 1B1 bands and the 2A1 and 
1B2 bands, respectively, had similar bandpasses, we used their weighted average 
for our further calculations. A point source in the aperture will result in a 
Gaussian with a width of about 18 pixels while a diffuse aperture filling 
source will yield a Gaussian with a width of 30 pixels. We have used this width 
to ensure that the signal in our observations was indeed of diffuse origin.

We have additionally searched the Digital Sky Survey plates from CDS\footnote{Centre 
de Donnes astronomiques de Strasbourg : http://cdsweb.u-strasbg.fr/} and 
found no point sources in the aperture. However, it is interesting to calculate 
the brightness of a star whose contribution would be equivalent to a 
diffuse flux of 20,000~\phunit. A star with a spectral type of later than about 
B9 would simply not have enough flux to contribute in the \fuse\ range without 
being blazingly bright in the visible. On the other hand, this amount of 
diffuse flux corresponds to an unreddened 18$^{th}$ magnitude B3 star implying 
a spectroscopic distance of about 1.5 kpc, or well beyond the Coalsack Nebula 
which would, of course, absorb any UV component of such a star.

\begin{figure}
\figurenum{1}
\plotone{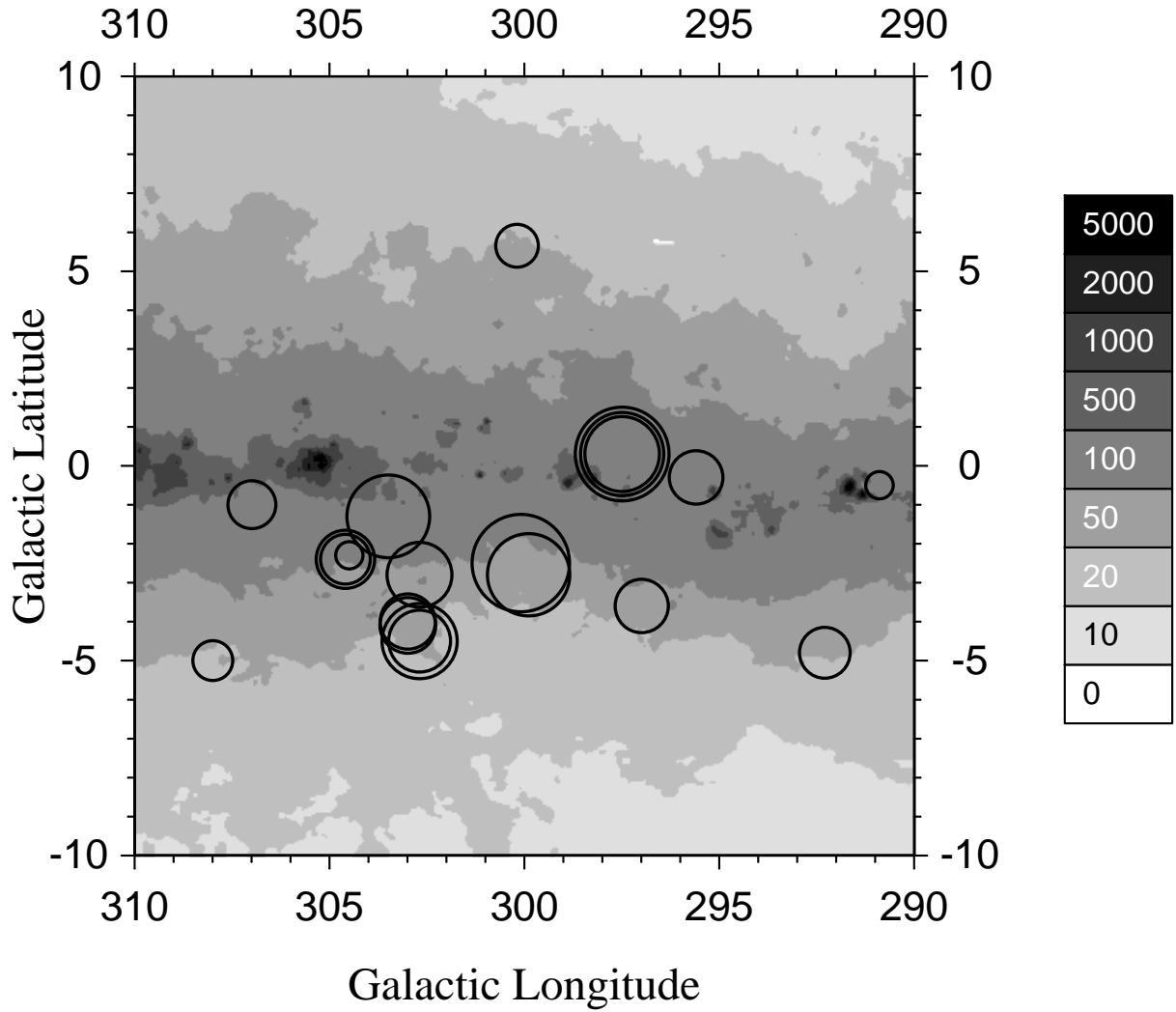}
\vspace{-4cm}
\figcaption{IRAS 100 $\micron$ (in units of MJy sr$^{-1}$) map of the region is
plotted with the observed locations marked as circles whose diameter is proportional 
to the weighted average intensity of the 2A2 and 1B1 bands (1114~\AA) in units 
of photons cm$^{-2}$ s$^{-1}$ sr$^{-1}$ \AA$^{-1}$.}
\label{locmap}
\end{figure}

Our observed values for each of the 6 \fuse\ bands and for the \voyager\ 
spectra at the same wavelengths are listed in Table \ref{voy_res} and are 
superimposed on a 100 \micron\ map from \iras\ in Fig. \ref{locmap}. The 
circles are centred on the observed locations and the diameter of each circle 
is proportional to the weighted average of the intensity in the 2A2 and 1B1 bands
at an effective wavelength of about 1114 \AA.

\section{RESULTS AND MODELING} 

\begin{figure}
\figurenum{2}
\epsscale{.95}
\plotone{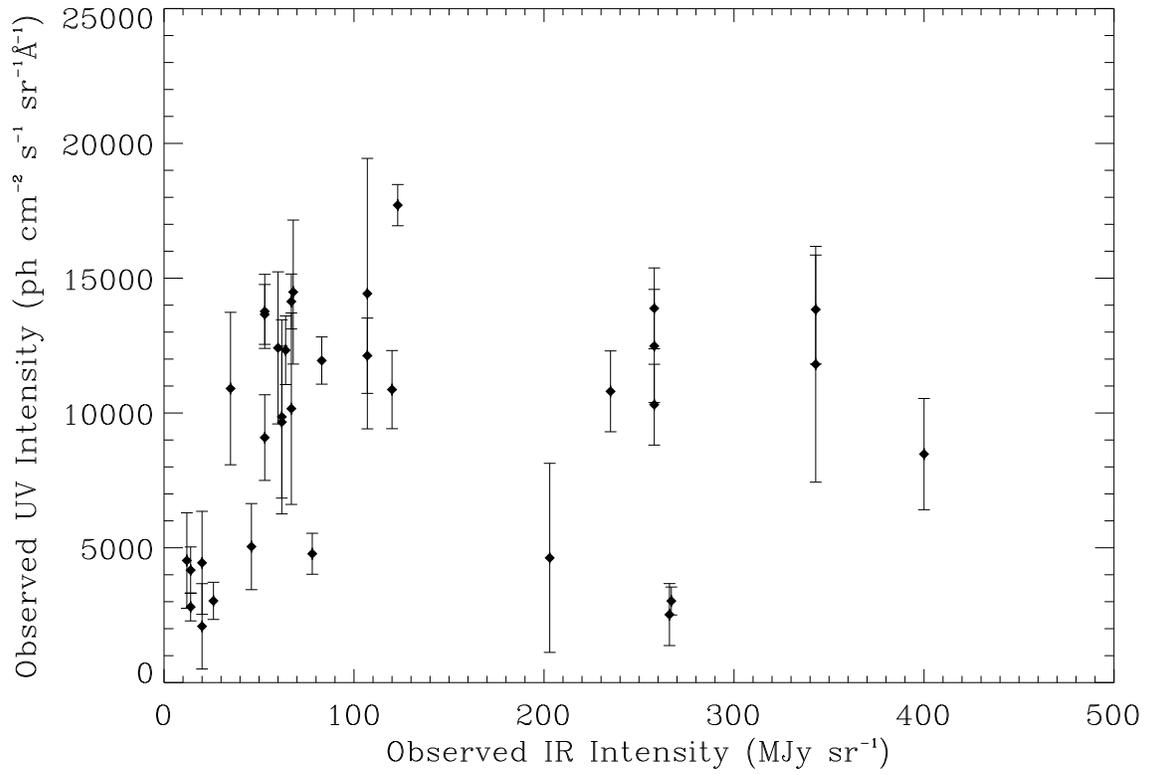}
\figcaption{Weighted average UV intensities of 2A2 and 1B1 bands (1114~\AA) with 
1$\sigma$ error bars are plotted against the observed IRAS 100 $\micron$ 
intensities at each location.}
\label{uvirobs}
\end{figure}

It is apparent from Fig. \ref{locmap} and Fig. \ref{uvirobs}, where the weighted 
average of the 2A2 and 1B1 bands are plotted against the 100~\micron\ intensity, 
that there is not a simple correlation between the UV and IR emission. There is 
a tendency for the amount of diffuse UV light to increase with the IR emission 
up to an intensity of about 80 MJy sr$^{-1}$, but with a lot of scatter. This 
correlation breaks down for larger IR intensities possibly suggesting that both 
the IR and UV emission are dominated by emission from the foreground cloud at 
lower H~{\small I} column densities (as traced by the IR) but not at higher column 
densities where the IR emission is largely due to emission from dust in the 
optically thick Coalsack molecular cloud. 

The scattered UV light from any location in space is a function of the 
interstellar radiation field (ISRF), the amount of dust in the line of sight and 
the scattering function of the dust grains. Of these, the ISRF is the easiest 
to derive as the Coalsack is so thick that no stars will be seen from behind 
the cloud, particularly in the UV, and the radiation field is dominated by only 
13 stars (Table \ref{star_inf}). As described by \citet{Suj04} we have used the 
Hipparcos catalog to locate the stars in 3-dimensions and calculated their 
contribution at the location of scattering based on their spectral type, V 
magnitude, and appropriate Kurucz models \citep{kurucz}. Not less than 95\% of 
the total ISRF in the vicinity of the Coalsack comes from these stars. This 
method is identical to that of \citet{Sh04} except that they had incorrectly 
scaled the FUV 
fluxes of the stars to observations made with the small aperture of the 
{\it International Ultraviolet Explorer}. The small aperture of {\it IUE} is 
known to underestimate stellar fluxes by about 40\% and thus they derived an 
albedo that was too high by the same factor. The FUV fluxes used in this work 
are in agreement with large aperture {\it IUE} observations of the stars.

The dust distribution has been well characterized by \citet{Co04} using 4 color 
photometry of several hundred stars in the region. They have found, in addition 
to the Coalsack Nebula itself at a distance of 180 pc, two foreground clouds of 
neutral hydrogen at distances of 60 pc and 120 - 150 pc. The column densities 
(N(H~{\small I})) of these clouds are 3.2 $\times$ 10$^{19}$~cm$^{-2}$ and 1.5 
$\times$ 10$^{21}$~cm$^{-2}$, respectively. We have used all three clouds in our 
modeling but note that most of the observed light comes from the more distant of 
the two H {\small I} clouds.

We have implemented a Monte Carlo code to account for multiple scattering in all 
three clouds: the two foreground neutral hydrogen clouds and the Coalsack molecular 
cloud. In this code, a photon is emitted in a random direction from one of the 
stars and continues in a straight line until it has an interaction with
a dust grain, the probability of which depends on the local density and the 
grain cross-section, taken from the ``Milky Way'' model of \citet{We01}.
This model uses a mixture of silicate and graphite grains with implicit 
assumptions of R$_{V}$ = 3.1 and the canonical gas-to-dust ratio of \citet{Bohlin}. 
After each interaction, the relative weight of the photon is reduced by 
the albedo and it is scattered into a new direction with a probability taken 
from the Henyey-Greenstein scattering function \citep{H.G}. Each individual 
photon is followed either until its weight becomes negligible or the 
photon escapes the region of interest. A complete run consists of about $10^{7}$ 
photons emitted for each star for each value of {\it $a$} and~{\it $g$}.

\begin{figure}
\figurenum{3}
\epsscale{.85}
\centering
\plotone{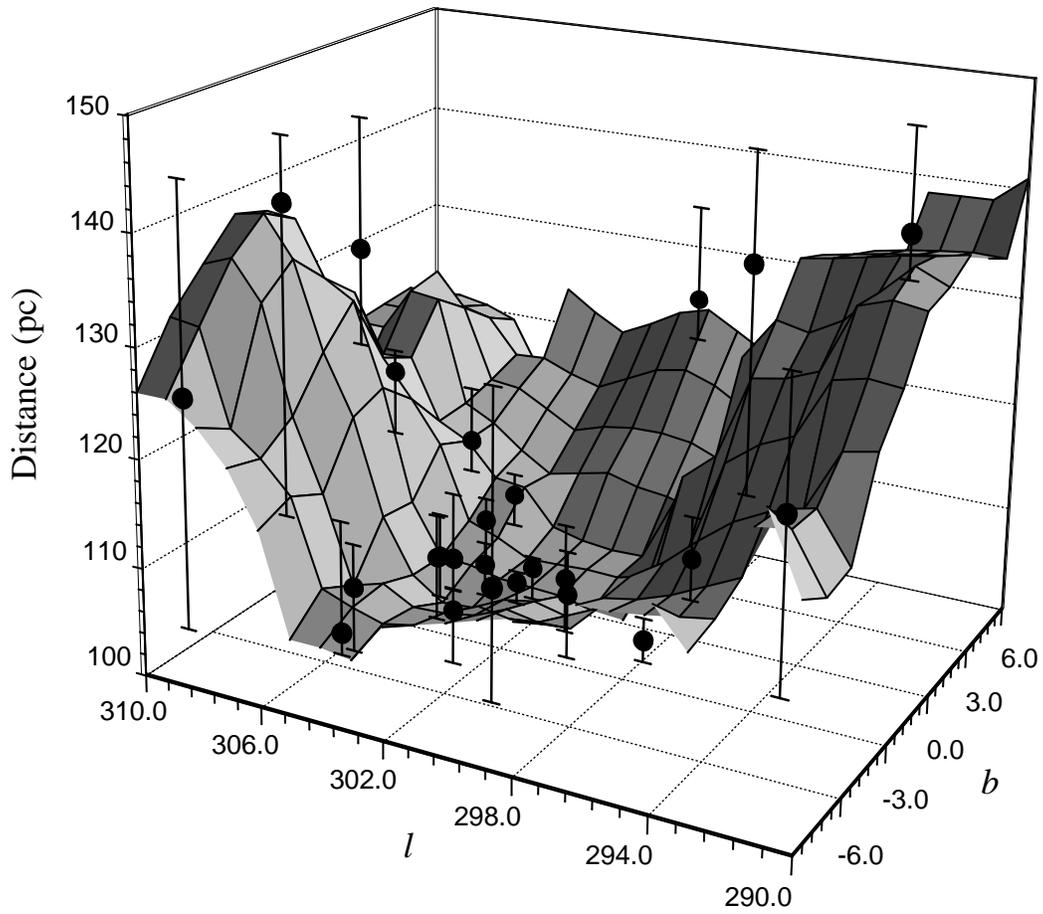}
\figcaption{Best fit distance of the more distant of the two H~{\small I} clouds 
(derived from the weighted average intensities of 2A2 and 1B1 bands at 1114~\AA, assuming 
that {\it $a$} and {\it $g$} remain constant throughout the region) is shown as dark 
circles with error bars showing the range of allowed distances. The interpolated surface 
fit for the region is also overplotted.}
\label{bestfit}
\end{figure}

We found that most of the observed radiation arose in the more distant of the 
two foreground clouds and hence most of the uncertainty in our model results 
comes from the uncertainty in the actual distance of that cloud. Because there 
is no reason to assume that the cloud is flat and perpendicular to our line of 
sight, we have derived the distance at each scattering location by finding the 
combination of optical constants ({\it $a$} and {\it $g$}) and distance which 
gives the best match of the predicted light with the observed value (weighted 
average of the 2A2 and 1B1 bands, i.e., at 1114~\AA), with the further assumption 
that the optical constants are the same throughout the region. These distances 
are plotted in Fig. \ref{bestfit} with error bars showing the range of allowed 
distances. Any point outside this allowed region will not satisfy our conditions 
of uniform {\it $a$} and {\it $g$}. Given the sparse nature of our data, we find 
a contiguous but warped cloud.

\begin{figure}
\figurenum{4}
\epsscale{1.0}
\centering
\plotone{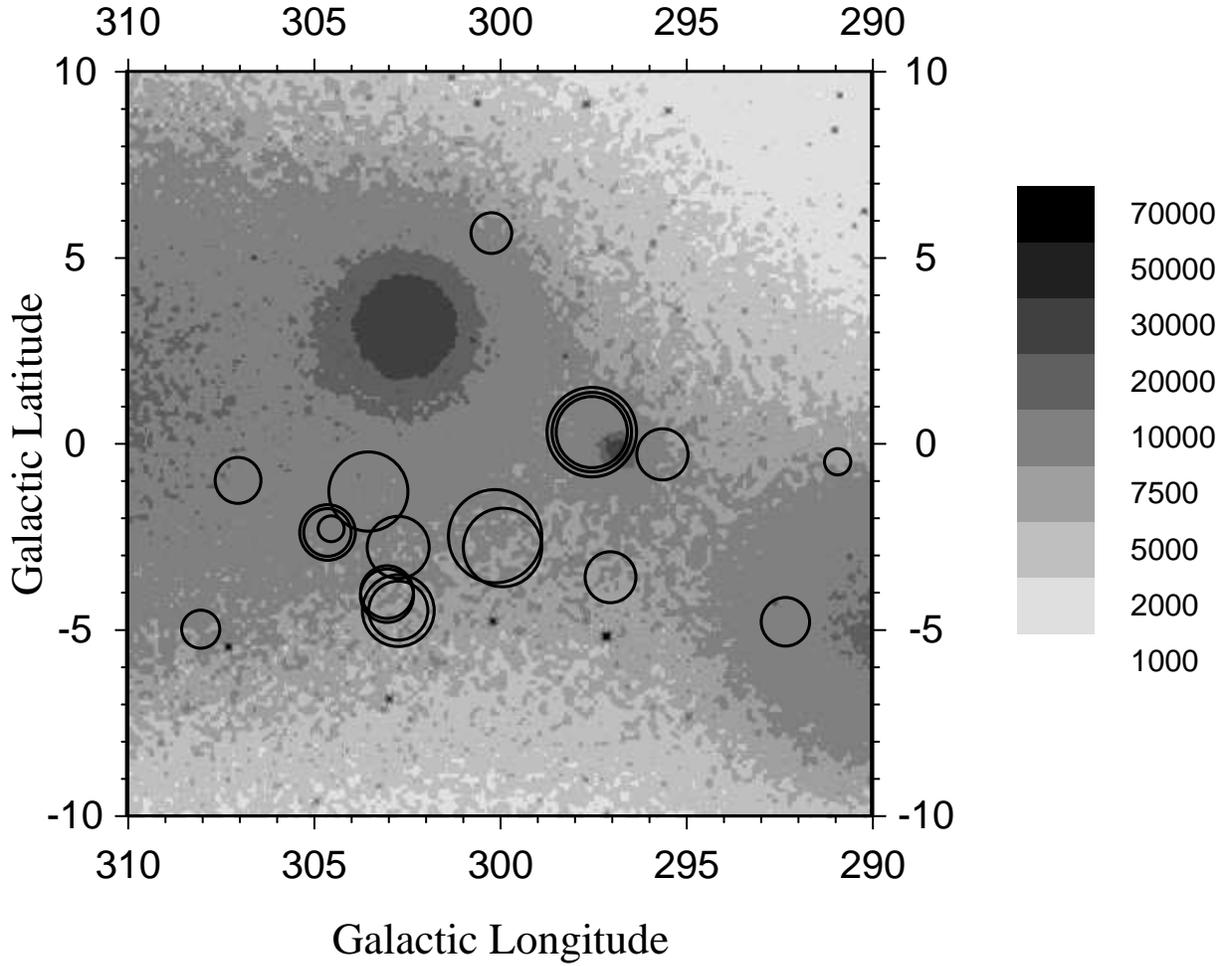}
\figcaption{The scattered light predicted by our model with $\it{a}$ = 0.28 and 
$\it{g}$ = 0.61 is shown in figure in units of photons cm$^{-2}$ s$^{-1}$ sr$^{-1}$ \AA$^{-1}$. The observed locations are overplotted as circles whose radii are proportional to their 
intensity at 1114~\AA.}
\label{uvint}
\end{figure}

Our final model assumes three clouds each with a 1 pc thickness (defined by our 
bin size): the Coalsack molecular cloud at a distance of 180 pc, a cloud of 
neutral hydrogen at a distance of 60 pc from the Sun, and the cloud illustrated 
in Fig. \ref{bestfit}b with a distance at each point as found from the best fit 
to the data. The output of this model is an image of the region around the 
Coalsack for each value of the optical constants which can be directly compared 
to the observations in each of the wavelength bands. Fig. \ref{uvint} shows this 
image for the best fit values of {\it $a$} and {\it $g$} (0.28 and 0.61, 
respectively) at a wavelength 1114~\AA, with our observations plotted as circles 
whose diameters are proportional to the weighted average of the 2A2 and 1B1 bands. 

The 6 \fuse\ bands (Table \ref{bands}) allowed observations at 4 wavelengths (1004~\AA,
1058~\AA, 1114~\AA, and 1158~\AA) where the intensities at 1114~\AA\ and 1158~\AA\ were 
taken from the weighted average of the 2A2 and 1B1 bands and 2A1 and 1B2 bands, 
respectively. The \voyager\ UVS is far more sensitive to diffuse radiation because 
of its relatively large aperture and allowed observation of the entire spectrum of the 
diffuse radiation between 912~\AA\ (the Lyman limit) and 1200~\AA.

\begin{figure}
\figurenum{5}
\epsscale{.95}
\plotone{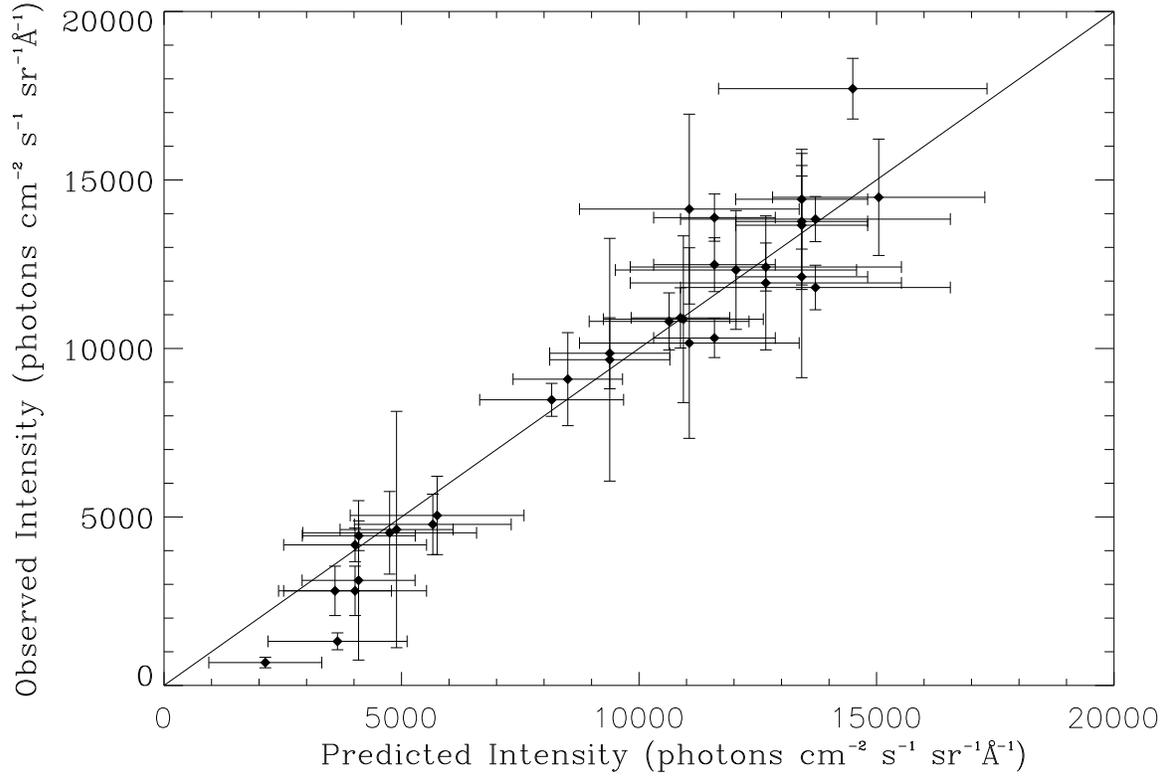}
\figcaption{The weighted average intensities of the 2A2 and 1B1 bands (1114~\AA) 
have been plotted against the predicted UV intensities at 1114~\AA\ with $\it{a}$ 
= 0.28 and $\it{g}$ = 0.61. The vertical error bars represent observational errors 
while the horizontal error bars represent model uncertainties.}
\label{obsmod}
\end{figure}

\begin{figure}
\figurenum{6}
\epsscale{.95}
\plotone{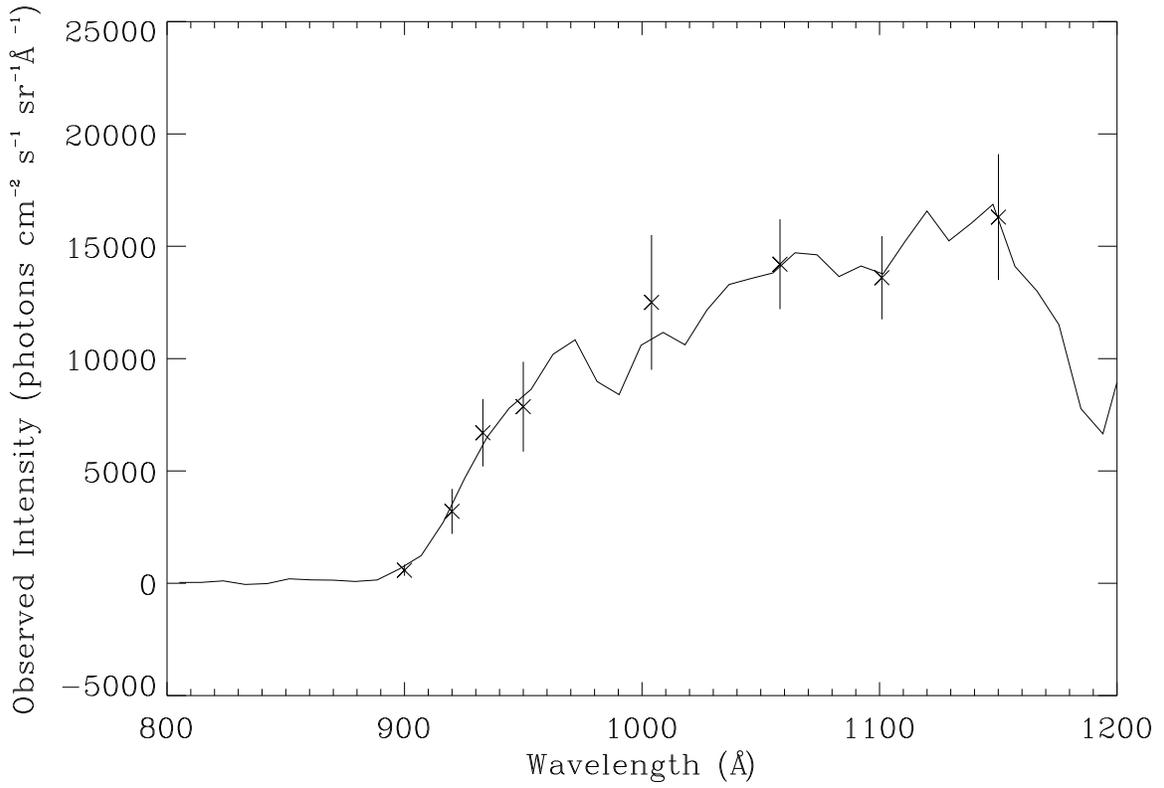}
\figcaption{Predicted intensities corresponding to the best fit parameters are shown as 
stars on a {\it Voyager} observation (No. 2  in Table \ref{voy_res}). The error bars 
correspond to the range allowed by the uncertainty in the optical constants.}
\label{vgrspe}
\end{figure}

\begin{figure}
\figurenum{7}
\centering
\plotone{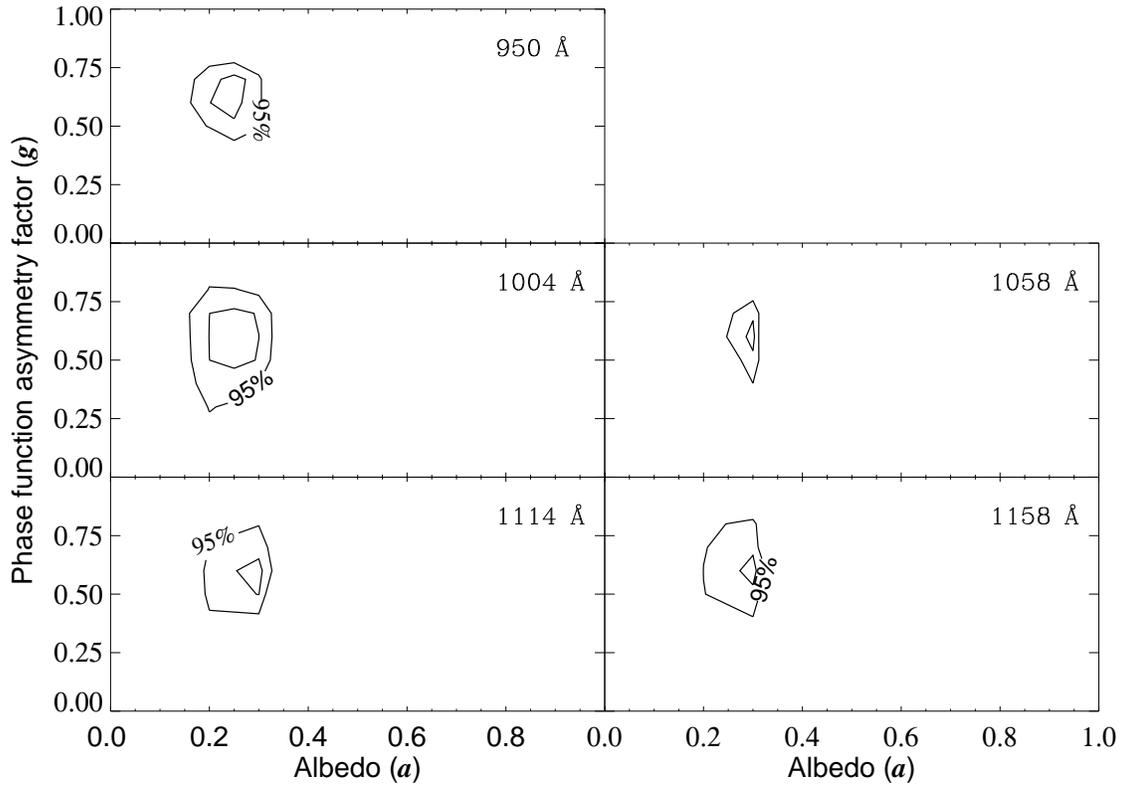}
\figcaption{67\% and 95\% confidence contours ($\it{g}$ versus $\it{a}$)  are
plotted for wavelengths 950, 1004, 1058, 1114 \& 1158~\AA. Only the 5 \voyager\ 
observations could be used to constrain the derived values at 950~\AA.}
\label{agcntr}
\end{figure}

\begin{figure}
\figurenum{8}
\epsscale{.9}
\plotone{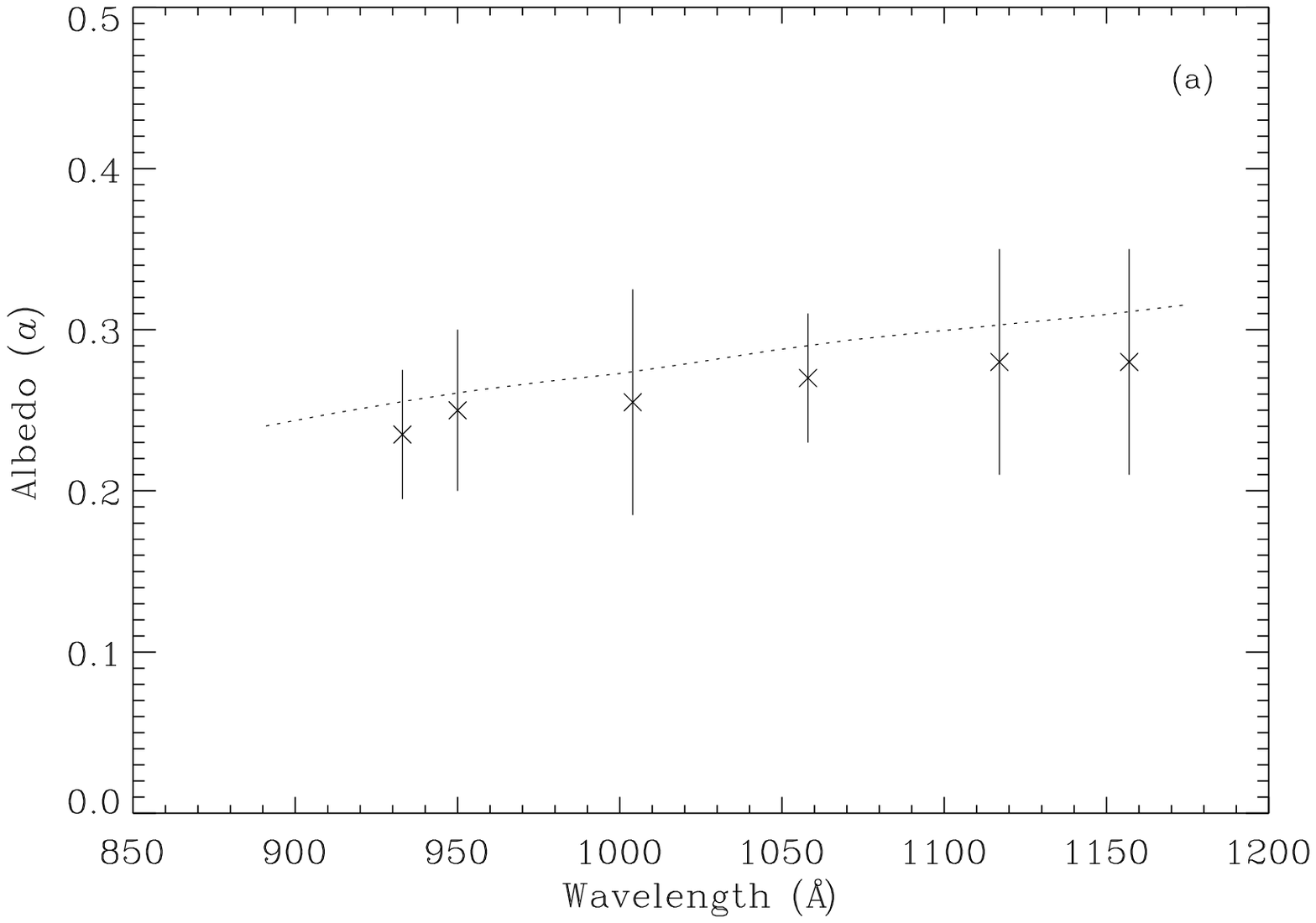}
\plotone{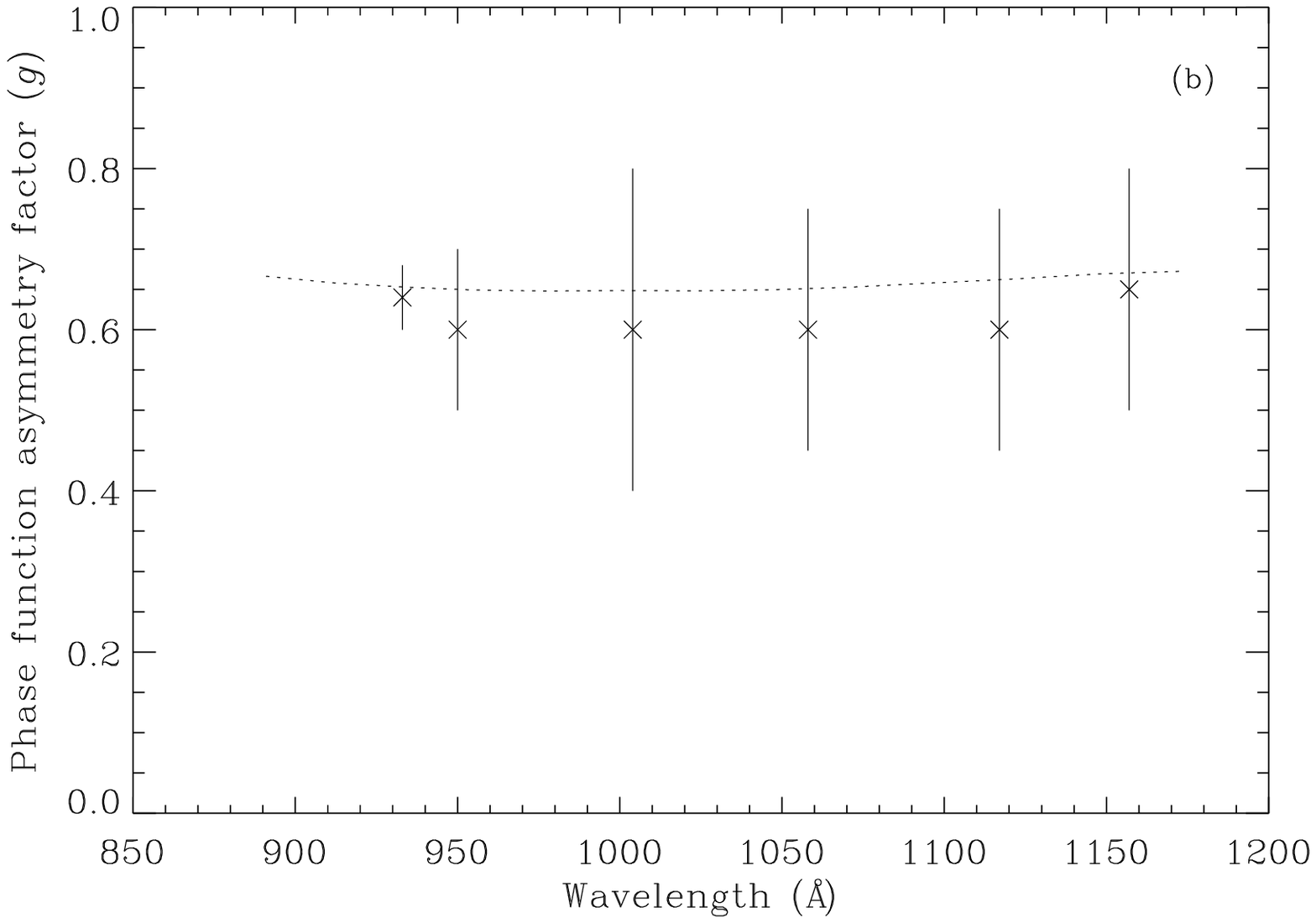}
\figcaption{The spectral variation in the albedo $\it{a}$ and in the phase function 
asymmetry factor $\it{g}$ are plotted in (a) and (b), respectively. The theoretical prediction  
of \citet{We01} is overplotted as dotted line. }
\label{ag_wave}
\end{figure}

Our predictions from our best fit model agree well with the observations both 
spatially (Fig. \ref{obsmod}) and spectrally (Fig.~\ref{vgrspe}). We have 
plotted 67\% and 95\% confidence contours (following the procedure of \citet{Lampton}) 
for {\it $a$} and {\it $g$} in Fig. \ref{agcntr}. They are consistent with values of 
$0.28 \pm 0.04$ for the albedo and $0.61 \pm 0.07$ for the phase function asymmetry 
factor throughout the spectral range from 912~\AA\ to 1200~\AA\ (Fig. \ref{ag_wave}), 
in agreement with the prediction of \citet{We01} for their ``Milky Way'' model. 
The error bars in the optical constants include both observational errors and errors
 in the modeling, such as in the distance.
\section{CONCLUSIONS}
We have used \voyager\ and \fuse\ observations of diffuse emission near the 
Coalsack Nebula to constrain the optical parameters of the interstellar dust. 
We find that the albedo {\it $a$} is $0.28 \pm 0.04$~and {\it $g$} is 
$0.61 \pm 0.07$~throughout the spectral range from 900 to 1200~\AA. 
These values are consistent with previous determinations in reflection Nebulae 
\citep{Witt93,Burgh02}, in diffuse clouds \citep{Suj05}, and in Orion 
\citep{Sh06}. It is clear that interstellar grains in the FUV are strongly forward 
scattering with a moderately low albedo, in agreement with theoretical prediction for 
a mixture of graphite and silicate grains \citep{We01}. Even though small grains have 
been depleted in Orion (R$_{V} = 5.5$; \citet{Fitzpatrick99}), it makes little difference 
to the optical constants \citep{We01} and our data cannot distinguish between them.

It had been our hope that we could derive a global model for the diffuse UV 
radiation over the entire sky. However, we have found the true situation to be 
more complex with the radiation being dependent largely on the presence of 
scattering dust near a hot star. In particular, we note that the {\it SPEAR} 
data \citep{Edelstein} show strong enhancements in the diffuse emission in the 
Ophiuchus and Coalsack regions which one might have naively associated with the 
prominent molecular clouds in those regions. However, our detailed modeling 
(\citet{Suj05} and this paper, respectively) have shown that the emission is 
actually due to scattering from a much thinner foreground cloud. We plan to 
continue our characterization of the diffuse UV radiation field and its 
implications for the nature of the interstellar dust using \voyager, \fuse\ 
and {\it GALEX} ({\it Galaxy Evolution Explorer}) observations.

\acknowledgements
We thank an anonymous referee for constructive criticism which we hope has 
resulted in a better paper. We thank the \fuse\ team for much helpful 
information and discussion. This research has made use of NASA's Astrophysics 
Data System and the SIMBAD database operated at CDS, Strasbourg, France. The 
data presented in this paper were obtained from the Multimission Archive at the 
Space Telescope Science Institute (MAST). STScI is operated by the Association 
of Universities for Research in Astronomy, Inc. under NASA contract NAS5-26555. 
Support for MAST for non-HST data is provided by the NASA Office of Space 
Science via grant NAG5-7584 and by other grants and contracts.

\bibliography{ms}
\clearpage
\pagestyle{empty}
\begin{deluxetable}{c c c c c c c c c c c c}
\rotate
\tabletypesize{\scriptsize}
\tablecaption{OBSERVED LOCATIONS IN THE COALSACK}
\tablewidth{0pc}
\tablehead{
 \colhead{No.}  &  \colhead{Data ID}  &  \colhead{Target Name}  &  \colhead{$\it{l}$}  &  \colhead{$\it{b}$}  & \multicolumn{6}{c}{Observed UV Intensity $\pm$ Error (photons cm$^{-2}$ s$^{-1}$ sr$^{-1}$ \AA$^{-1}$)}  &  \colhead{IR 100 \micron} \\
   &   &   & \colhead{(deg)}  &  \colhead{(deg)}  &  \colhead{1A1}  &  \colhead{1A2}  &  \colhead{2A2}   &  \colhead{1B1}  &  \colhead{1B2} &  \colhead{2A1}  & \colhead{(MJy sr$^{-1}$)}\\
   &   &   &   &    &  \colhead{(1004~\AA)}  &  \colhead{(1058~\AA)}  &  \colhead{(1112~\AA)}   &  \colhead{(1117~\AA)}  &  \colhead{(1157~\AA)} &  \colhead{(1159~\AA)}  & \\
 }

\startdata
1& Voyager 1\tablenotemark{a}  & BKGND3   &301.7&-1.7&  13165 $\pm$ 366 & 16212 $\pm$ 590 &  17023 $\pm$ 730  & 18500 $\pm$ 800  & 23675 $\pm$ 1500  & 23700 $\pm$ 1500  &  123   \\[0.2cm]
2& Voyager 2\tablenotemark{b}  & Coalsack   &303.7&0.8& 9240 $\pm$ 1000  & 10750  $\pm$ 545 &  13950 $\pm$ 2000   & 11519 $\pm$ 700  & 15120  $\pm$ 800  & 15210  $\pm$ 1000 &  343  \\[0.2cm]
3& Voyager 3\tablenotemark{b}   & Coalsack   &303.7&0.8& 10880 $\pm$ 700  & 13815  $\pm$ 700  &  14000 $\pm$ 2300   & 13823  $\pm$ 700  & 15916  $\pm$ 1200  & 14104  $\pm$ 800  & 343  \\[0.2cm]
4& Voyager 4\tablenotemark{b}   & Coalsack  &304.6&-0.4& 4311 $\pm$ 500   & 6140  $\pm$ 500  &  11900 $\pm$ 2400   & 8295 $\pm$ 500  & 11150 $\pm$ 800  & 11000 $\pm$ 1000  &  400   \\[0.2cm]
5& Voyager 5\tablenotemark{b}   & Coalsack   &305.2&-5.7& 9450 $\pm$ 500  & 11060 $\pm$ 700  &  8000 $\pm$ 2000    & 11520 $\pm$ 1000  & 16720 $\pm$ 1200  & 15500 $\pm$ 1500  &  35  \\[0.2cm]
6&  B0680101  &  Gamma-Cru  &300.17&5.65&  1045 $ \pm$ 792  &  3077 $ \pm$ 825  &  3031 $ \pm$ 518  &  289 $ \pm$ 219  &  539      $ \pm$ 408  &  3818 $ \pm$ 938  &  26     \\[0.2cm]
7& D0260101 & HD113708 &304.55&-2.39& 8830   $ \pm$  2995 & 9228  $\pm$ 2450     & 6640 $\pm$   2010     &        20065 $\pm$ 2188  & 20591  $\pm$ 2444 & 6780  $\pm$ 2126 & 107    \\[0.2cm]
8& D0260102 & HD113708 &304.55&-2.39& 5304 $\pm$ 3078 & 7611  $\pm$ 2093   & 6627  $\pm$  5022     &  14273 $\pm$  3724   & 11357  $\pm$ 2194 & 5788  $ \pm$ 4386 & 107    \\[0.2cm]
9& D0260201 & HD113659 &304.52&-2.26& 7544   $ \pm$  4087 & 7254  $ \pm$ 1815     & 6468 $\pm$  3903     &        13014 $\pm$  3199  & 10914 $\pm$ 1836 & 4074 $\pm$ 3087 & 120    \\[0.2cm]
10& D0260301 & HD111641 &302.97&-3.98& 3647  $ \pm$  2339 & 6132  $ \pm$ 1319     & 3031        $ \pm$  518      &        13779   $ \pm$  2020     & 10584 $ \pm$    1736  & 3461 $ \pm$ 674 & 53     \\[0.2cm]
11& D0260302 & HD111641 &302.97&-3.98& 8422  $ \pm$  2586 & 8733  $ \pm$ 1547     & 12287      $ \pm$  3606     &        14065   $ \pm$  2035     & 13242 $ \pm$    1723 & 4803 $ \pm$ 2941 & 53     \\[0.2cm]
12& D0260401 & HD111195 &302.65&-4.49& 5772  $ \pm$  2082 & 8091  $ \pm$ 1472     & 8838       $ \pm$  1693     &        10249   $ \pm$  1482     & 10459 $ \pm$    1473 & 7687 $ \pm$ 1814 & 62     \\[0.2cm]
13& D0260402 & HD111195 &302.65&-4.49& 8194  $ \pm$  2188 & 10160  $ \pm$        1561     &        10778   $ \pm$  1576     &        9044    $ \pm$  1419 &    8583 $ \pm$      1990 & 11029 $ \pm$        1790     & 62     \\[0.2cm]
14& D0260501 & HD111283 &302.69&-2.72& 6666  $ \pm$  4015 & 9077  $ \pm$ 2191     & 6648       $ \pm$  5038     &        12683   $ \pm$  2164     & 15342 $ \pm$    3167 & 7048 $ \pm$ 4859 & 83     \\[0.2cm]
15& D0260601 & HD116796 &306.94&-0.95& 4338  $ \pm$  2510 & 4139  $ \pm$ 1650     & 4627       $ \pm$  3506     &  4827 $ \pm$ 897   &  3119 $ \pm$ 2364 & 3791 $ \pm$ 2708 & 203    \\[0.2cm]
16& D0260701 & HD117667 &299.95&-2.73& 23614 $ \pm$  5031 & 22132      $ \pm$  4400     &        16511   $ \pm$  6952     & 13621  $ \pm$  3080     & 14986 $ \pm$    4366 & 12930 $ \pm$ 6116     & 67     \\[0.2cm]
17& D0260702 & HD117667 &299.95&-2.73& 9626  $ \pm$  4093 & 12149      $ \pm$  2720     & \nodata\tablenotemark{c}  & 10161  $ \pm$  2827     & 4667 $ \pm$     3537 &  \nodata\tablenotemark{c}   & 67   \\[0.2cm]
18& E0290101 & Coalsack-1 &303.52&-1.32& 8926        $ \pm$  1725     & 11212  $ \pm$  824      &        10515   $ \pm$  1375     &        10976   $ \pm$  1073     & 9025   $ \pm$  999 & 7224        $ \pm$  1637     & 235    \\[0.2cm]
19& E0290301 & Coalsack-3 &297.02&-3.62& 3678        $ \pm$  2787     & 5685   $ \pm$  1544     &        5389    $ \pm$  3621     &        13792   $ \pm$  2019     & 13043  $ \pm$  1994 & 3147 $ \pm$ 2385     & 64     \\[0.2cm]
20& E0290401 & Coalsack-4 &308.01&-4.99& 4216        $ \pm$  2278     & 5957   $ \pm$  1067     &        3741    $ \pm$  2163     &        5487    $ \pm$  1379     & 4861   $ \pm$  1544 & 3024 $ \pm$ 1669     & 46     \\[0.2cm]
 \tableline
 \tablebreak
21& S4050701 & HD96548-BKG &292.32&-4.83& 7051       $ \pm$  766      & 9408   $ \pm$  1521     &        8270    $ \pm$  2461     &        9441    $ \pm$  1667     & 8141   $ \pm$  1284 & 8979 $ \pm$ 1148     & 53     \\[0.2cm]
22& S4051701 & HD104994-BKGD &297.56&0.34& 10005  $ \pm$  1305     &        12378   $ \pm$  796      & 17134  $ \pm$  1162     & 11626  $ \pm$  878      & 11241 $ \pm$    980      &        11823   $ \pm$  1318     & 258    \\[0.2cm]
23& S4055301 & WR42-HD97152-BGD &290.95&-0.49& 660 $ \pm$ 500    & 1288   $ \pm$  541      &        3031    $ \pm$  518      &        144     $ \pm$  109      &        360 $ \pm$ 273     & 3461   $ \pm$  674      & 267    \\[0.2cm]
24&  S4055801  &  HD102567-BKGD  &295.61&-0.24&  3538  $ \pm$  1256  &  5279 $ \pm$ 455  &  1767 $ \pm$ 527      &  3452 $ \pm$      690      &  3206  $ \pm$  849  &  5330 $ \pm$        915  &  266   \\[0.2cm]
25& S4059101 & HD104994-BKGD &297.56&0.34& 8711   $ \pm$  1056 & 10994 $ \pm$        589      &        11305   $ \pm$  1059     &        9852    $ \pm$  693      & 8971   $ \pm$  618 & 9659 $ \pm$  715      & 258    \\[0.2cm]
26& S5052801 & HD108002-BKGD &300.16&-2.48& 10808  $ \pm$  3230 & 17048 $ \pm$        1309 & 16862 $ \pm$      3320     & 13498 $ \pm$ 2022   & 11437  $ \pm$  2069 & 13198 $ \pm$ 3158 & 68     \\[0.2cm]
27& S5059001 & POLE-BKGD &307.12&-2.44& 1953   $ \pm$  1359 & 5221 $ \pm$ 682      & 7151  $ \pm$  2213     & 4157   $ \pm$  981 & 6476 $ \pm$ 868 & 2498 $ \pm$  1317 & 78     \\[0.2cm]
28& S5059101 & POLE-BKGD &303.9&-8.14& 1458   $ \pm$  724 & 4522 $ \pm$  565      & 5985 $ \pm$  1032     & 4019   $ \pm$  488 & 3646  $ \pm$  927 & 2136 $ \pm$  1042 & 20     \\[0.2cm]
29& S5059102 & POLE-BKGD &303.9&-8.14& 1707   $ \pm$  1294 & 4872 $ \pm$ 1622     & 2044   $ \pm$  1549     & 2134   $ \pm$  1617 & 1985 $ \pm$  1304 & 2213 $ \pm$ 1677 & 20     \\[0.2cm]
30& S5059201 & POLE-BKGD &301.97&-2.14& 10641  $ \pm$  1674 & 13708 $ \pm$        845 &        16677   $ \pm$  1393     &     10501   $ \pm$  829      & 9378   $ \pm$  924 & 8117 $ \pm$  1940     & 60     \\[0.2cm]
31& S5059302 & POLE-BKGD &298.92&-8.51& 2792   $ \pm$  1513 & 4377 $ \pm$ 1359     & 4769   $ \pm$  2393     & 4441   $ \pm$  1428 & 2767 $ \pm$  1501 & 1645 $ \pm$ 849      & 12     \\[0.2cm]
32& S5160101 & HD104994 &297.56&0.34& 8475   $ \pm$  1395 & 12669 $ \pm$        695 &  15611   $ \pm$  1863     &     11667   $ \pm$  884      & 10333  $ \pm$  931 & 12731 $ \pm$  1245     & 258    \\[0.2cm]
33& S5058901 & POLE-BKGD &308.54&-8.86& 645    $ \pm$  489 & 2220 $ \pm$  515      & 4948   $ \pm$  1169     & 3980   $ \pm$  550 & 3918 $ \pm$ 652 & 4156 $\pm$  928  & 14     \\[0.2cm]
34& S5058902 & POLE-BKGD &308.54&-8.86& 1155   $ \pm$  875 & 2001 $ \pm$  710      & 1958 $ \pm$1448 & 3046   $ \pm$  851 & 2699  $ \pm$  978 & 1118 $ \pm$  847      & 14      \\[0.2cm]
\enddata
\label{voy_res}
\tablenotetext{a}{\citet{Mu99}}
\tablenotetext{b}{\citet{Mu94}}
\tablenotetext{c}{Data nonexistent}
\end{deluxetable}
\clearpage
\pagestyle{plaintop}
\begin{deluxetable}{c c c c}
\tabletypesize{\small}
\tablecaption{\fuse\ \ WAVELENGTH BANDS}
\tablewidth{0pt}
\tablehead
{\multicolumn{2}{c}{Detector bands} & \colhead{Wavelength range} & \colhead{Average Wavelength} \\
& & \colhead{(\AA)} & \colhead{(\AA)}
}
\startdata
LiF & 1A1 & 987.1 - 1020.8   & 1004  \\
LiF & 1A2 & 1034.8 - 1081.4  & 1058 \\
LiF & 2A2 & 1095.0 - 1128.6 & 1112  \\
LiF & (2A2+1B1)/2\tablenotemark{*} &  & 1114  \\
LiF & 1B1 & 1100.3 - 1133.7  & 1117 \\
LiF & 1B2 & 1133.7 - 1180.1  & 1157 \\
LiF & (1B2+2A1)/2\tablenotemark{*}  &  & 1158 \\
LiF & 2A1 & 1142.0 - 1175.3 & 1159   \\
\enddata
\label{bands}
\tablenotetext{*}{Derived band}
\end{deluxetable}

\begin{deluxetable}{c c c c c c c}
\tabletypesize{\scriptsize}
\tablecaption{BRIGHTEST STARS IN THE REGION}
\tablewidth{0pt}
\tablehead
{\colhead{HD Number} &\colhead{Name}&\colhead{\it{l}}&\colhead{\it{b}}& \colhead{Sp. Type\tablenotemark{a}} &\colhead{Distance\tablenotemark{a}}&\colhead{Luminosity\tablenotemark{b} at 1100~\AA} \\
        &    & \colhead{(deg)} &\colhead{ (deg)} &  &\colhead{ (pc)} & \colhead{(photons s$^{-1}$ \AA$^{-1}$)}
}
\startdata
122451& $\beta$ Cen   & 311.77 & 1.25 & B1III  & 161.3 & 2.45$\times$10$^{46}$\\
108248& $\alpha$ Cru  & 300.13 & -0.36& B0.5IV & 98.3  & 1.28$\times$10$^{46}$\\
111123& $\beta$ Cru   & 302.46 & 3.18 & B0.5IV & 108.1 & 1.0$\times$10$^{46}$\\
93030 & $\theta$ Car  & 289.6  & -4.9 & B0Vp   & 134.6& 4.62$\times$10$^{45}$\\
104841&$\theta$ Cru & 297.64 &-0.78 & B2IV   & 230.9& 1.38$\times$10$^{45}$\\
99264 &   & 296.32 &-10.51& B2IV-V & 271.0& 1.17$\times$10$^{45}$\\
91465 &PP Car & 287.18 &-3.15 & B4Vne  & 152.4& 6.87$\times$10$^{44}$\\
102776&J Cen   & 296.18 & -1.73& B3V    & 140.9& 3.45$\times$10$^{44}$\\
92938 &V518 Car & 289.56 & -5.00& B3V    & 139.9& 2.29$\times$10$^{44}$\\
93607 &  & 289.97 & -4.69& B3IV   & 137.7& 1.95$\times$10$^{44}$\\
103884&Glazar Cru 135& 296.76 & -0.22& B3V    & 183.5& 1.77$\times$10$^{44}$\\
93194 &  & 289.50 & -4.46& B5Vn   & 148.4& 6.61$\times$10$^{43}$\\
99103 &  & 293.78 & -3.66& B5    & 145.6& 6.00$\times$10$^{43}$\\
\enddata
\label{star_inf}
\tablecomments{ Stars in descending order of UV luminosity}
\tablenotetext{a}{From Hipparcos Catalog \citep{Hipparcos}}
\tablenotetext{b}{Using Kurucz Model scaled to V magnitude.}
\end{deluxetable}

\clearpage

\clearpage
\thispagestyle{empty}
\setlength{\voffset}{-26mm}
\clearpage
\setlength{\voffset}{0mm}


\end{document}